# Artificial Intelligence, Domain AI Readiness, and Firm Productivity

Sipeng Zeng  
CKGSB

Xiaoning Wang  
UT Dallas

Tianshu Sun  
CKGSB

## Abstract

Although Artificial Intelligence (AI) holds great promise for enhancing innovation and productivity, many firms struggle to realize its benefits. We investigate why some firms and industries succeed with AI while others do not, focusing on the degree to which an industrial domain is technologically integrated with AI, which we term "domain AI readiness". Using panel data on Chinese listed firms from 2016 to 2022, we examine how the interaction between firm-level AI capabilities and domain AI readiness affects firm performance. We create novel constructs from patent data and measure the domain AI readiness of a specific domain by analyzing the co-occurrence of four-digit International Patent Classification (IPC4) codes related to AI with the specific domain across all patents in that domain. Our findings reveal a strong complementarity: AI capabilities yield greater productivity and innovation gains when deployed in domains with higher AI readiness, whereas benefits are limited in domains that are technologically unprepared or already obsolete. These results remain robust when using local AI policy initiatives as instrumental variables. Further analysis shows that this complementarity is driven by external advances in domain-AI integration, rather than firms' own strategic pivots. Time-series analysis of IPC4 co-occurrence patterns further suggests that improvements in domain AI readiness stem primarily from the academic advancements of AI in specific domains.

**Keywords:** Artificial Intelligence, Readiness, Productivity, Complementarity, Patents

# 1. Introduction

In recent years, Artificial Intelligence (AI) has become one of the most transformative technologies reshaping the business landscape, fueled by recent progress in machine learning algorithms, data availability, and computing power. An increasing number of firms across various industries have invested heavily in AI, aiming to improve operational efficiency, enhance decision-making, and drive innovation (Rock, 2019; Lou & Wu, 2021; Babina et al., 2024; Wu et al., 2025). The widespread adoption of AI reflects a broader belief that AI can serve as a critical source of competitive advantage in the digital economy.

However, despite substantial investments, a growing body of reports and academic studies has highlighted that many firms have struggled to generate meaningful returns from their AI initiatives, known as the "AI productivity paradox" (Brynjolfsson et al., 2019; Brynjolfsson et al., 2021; Fraisse & Laporte, 2022). For example, using U.S. census data, McElheran et al. (2025) show a 2% productivity loss after firms have adopted AI. Surveys of business executives often reveal frustration with the underperformance of AI projects (Schlegel et al., 2023; Copper 2024). This paradox between high expectations and modest realized benefits raises important questions about the conditions under which AI can truly create value for firms.

One key explanation for this paradox is that successful AI adoption requires more than technological investment; it depends critically on the presence of complementary organizational and environmental factors (Raisch & Krakowski 2021). While both AI and traditional information technologies (IT) are often considered general-purpose technologies (GPTs), successful AI adoption requires a distinct set of complementary conditions that go beyond technological investment. Prior research has emphasized internal organizational complements for IT—such as management practices, decentralized decision-making, and workforce reskilling (Boynton et al., 1994; Malone, 1997; Bresnahan et al.,2002; Boothby et al., 2010; Aral et al., 2012; Tambe et al., 2012). However, AI differs from traditional IT in important ways: it relies heavily on context-specific data, metaknowledge, and continuous feedback loops, necessitating deeper integration with core technological processes in a domain field (Berente et al., 2021; Fügener et al., 2021; Fügener et al., 2022; Tamble, 2025). As a result, the domain field's alignment with AI—reflected in the availability, compatibility, and metaknowledge of data, typically shaped by factors external to the firm—plays a far more nuanced and critical role in AI adoption.

When examining the external factors that complement traditional IT, prior research has primarily focused on macro-level geographic infrastructures—for instance, how internet bandwidth influences firms' returns to IT adoption (Forman et al., 2005; Racherla & Mandviwalla, 2013). This emphasis stems from the fact that IT value realization is often mediated by large-scale network connectivity and communication infrastructures that determine the efficiency of information exchange across regions. In contrast, AI differs in that its effectiveness relies more heavily on domain-specific data availability, task-level knowledge integration, and industry-specific technological complementarities (Wu et al., 2020; Brey & Marel, 2024; Tambe, 2025). As a result, the returns to AI adoption are more sensitive to the degree of technological alignment and integration between AI and the focal industrial domain. For instance, while IT systems were rapidly deployed in retail for inventory and point-of-sale management, AI-based recommendation and personalization models require years of accumulating industry- and context-specific data as well as iterative model refinement (Sun et al., 2024; Yuan et al., 2025). We term the degree to which an industrial domain is technologically integrated with AI as domain AI readiness—a critical but underexplored external complement that shapes firms' ability to translate AI capabilities into performance gains.

To investigate this issue, we construct a novel firm-year panel dataset based on Chinese listed companies' financial and patent information. We focus on emerging markets because the rapid yet uneven diffusion of digital technologies across different industry sectors provides a unique empirical setting to examine how domain readiness interacts with firm AI adoption, which is often shaped by policy-driven initiatives in these markets (Wang & Wu, 2024). Leveraging the co-occurrence of four-digit International Patent Classification (IPC4) classification codes in AI patents, we construct a time-varying measure of AI readiness for each domain and derive each firm's annual AI readiness by averaging the domain-level readiness across all technological fields represented in its patent portfolio. Firms' AI capabilities are comprehensively assessed through their AI-related strategic disclosures, AI-focused talent recruitment, and AI asset accumulation. We measure the financial and innovative performance of companies using revenue per employee, total factor productivity, and number of trademarks.

Our empirical analysis yields several key findings. First, we establish a strong positive correlation between firms' AI capabilities and the degree of AI readiness in their technological fields, satisfying the demand test in the complementarity framework. Second, performance tests reveal that AI capabilities and AI readiness exhibit a complementary relationship: firms achieve greater improvements in innovation output and productivity when both AI capability and AI readiness are high. Third, by disentangling the sources of variation in AI readiness, we show that the complementarity is primarily driven by advances in external domain-AI integration, rather than by firms' strategic pivots in domains. Instrumental variable analyses exploiting regional AI policy shocks further corroborate the robustness of our findings.

This study makes several theoretical contributions to the information systems (IS) literature. First, it extends research on the IT and AI productivity paradox by demonstrating that the external technological environment—particularly the degree of domain-specific technological integration—is a critical determinant of AI returns, a factor relatively underexplored compared to internal organizational complements or regional infrastructure conditions. Our results show that both premature and obsolete adoption yield limited benefits. Second, it advances the understanding of AI as a general-purpose technology by showing that, unlike traditional IT, whose value realization often depended on broad infrastructural complements, the effectiveness of AI adoption is more tightly bound to industry conditions, reflecting its reliance on domain-specific data, specialized expertise, iterative learning, and algorithmic adaptation. This perspective provides scholars and practitioners with a sharper lens for understanding and strategizing AI adoption. Third, our findings underscore the policy and practical implications of this distinction between AI and traditional IT. This is particularly salient for emerging economies with uneven industrial structures. While some sectors—such as fintech, mobile payments, e-commerce, and social media—are highly digitalized and data-rich, others, especially traditional industries, lag behind in digital transformation. Faced with such disparities, policymakers in emerging markets should pursue differentiated policy pathways that target industry-specific AI readiness, with a focus on data governance, talent development, intellectual property, and technological integration within domains. Overall, the study contributes to building a more nuanced and context-sensitive framework for understanding the conditions under which the return of nascent technology adoption is maximized.

## 2. Theory and Literature

### 2.1 The "AI Productivity Paradox"

AI has evolved from a speculative academic pursuit into a transformative GPT with widespread applications across industries. Over the past decade, advances in machine learning, natural language processing, and computer vision have dramatically expanded the potential for firms to leverage AI to enhance operational efficiency, foster innovation, and create new business models (Shepherd & Majchrzak, 2022; Gofman & Jin, 2024). As a result, firms across a broad range of sectors have aggressively invested in AI capabilities, seeking to capture competitive advantages in an increasingly digital economy.

Despite the high expectations surrounding AI, accumulating empirical evidence indicates that many firms struggle to realize tangible value from their AI investments. Reports from news outlets, consulting firms, and academic studies alike have documented the so-called "AI productivity paradox" (McElheran et al., 2025), wherein substantial AI investment fails to consistently translate into improved firm performance (Brynjolfsson et al., 2019; Brynjolfsson et al., 2021; Fraisse & Laporte, 2022). Several explanations have been proposed, including misaligned organizational structures, insufficient data infrastructure, and lack of specialized talent (Rock, 2019; Anthony et al., 2023; Hui et al., 2024). However, there remains little empirical evidence examining the mechanisms driving the gap between AI investment and realized value. This persistent puzzle suggests the need for further investigation into the conditions under which AI adoption contributes to firm-level performance gains, motivating a closer examination of potential complementary factors.

### 2.2 Complementarity as an Explaining Mechanism

One prominent theoretical lens for understanding the uneven returns to technology investment is the concept of complementarity. Originating from studies of traditional IT, research has consistently found that the value of IT investments depends critically on the presence of complementary organizational changes, such as process reengineering, decentralized decision-making, and employee reskilling (Bresnahan et al., 2002; Brynjolfsson & Hitt, 2000; Tambe et al., 2012). As digital technologies evolved, this perspective expanded to encompass big data analytics, where organizational data capabilities, analytical talent, and decision-making structures serve as vital complements to data-driven initiatives (Ghosh et al., 2019; Vidgen

et al., 2017).

In the context of AI, scholars have increasingly applied the complementarity framework to explain performance heterogeneity. An emerging stream of research has highlighted the importance of internal factors such as data governance, lean and agile management processes, and the upskilling of employees to complement AI investments (Li, 2023; Babina et al., 2024; Wang & Wu, 2024). However, while these studies focus on the internal organizational complements to AI adoption, much less attention has been given to the external technological environment—that is, the industry- or domain-specific factors that may condition the effectiveness of AI initiatives. This is because complementarity operates not only within firms but also between firms and their domains, where the degree of domain–AI integration may critically shape adoption outcomes. Understanding this external dimension requires engaging more directly with the technological nature of AI itself.

In this study, we align with recent literature defining AI as advanced computational power for generating predictive actions through machine learning techniques, often prioritizing accuracy over interpretability (Agrawal et al. 2018, Taddy 2018, Berente et al. 2021, Brynjolfsson et al. 2021, Wang & Wu 2024). The "black-box" nature of AI algorithms underscores the importance of contextualizing the data on which they are trained (Berente et al. 2021, Li et al. 2021, Taddy 2018). Building on this perspective, our study emphasizes that such contextualization is inherently domain-specific: the effectiveness of AI adoption critically depends on the extent to which industry environments provide the requisite data richness, domain expertise, and technological integration (Brynjolfsson et al. 2021, Tambe 2025). This focus highlights a fundamental distinction between AI and traditional IT, and explains why the returns to AI adoption vary more systematically with industry-level conditions than with broad infrastructural complements.

**2.3 Domain AI Readiness Complements AI Adoption**

Research on traditional IT adoption has recognized that external conditions—such as regional IT infrastructure, digital policy environments, and supply chain digitization—can serve as critical complements that influence the success of IT investments (Forman et al., 2005; Mithas et al., 2012). However, these external complements typically operated at a macro, geographic scale, focusing on regional differences rather than domain-specific technological characteristics. In contrast, the nature of AI as a GPT introduces a fundamentally different type of external complementarity. Unlike traditional IT systems, which can often be deployed with minimal adaptation across industries, AI deployment is highly sensitive to the technological characteristics of the target domain. Effective AI applications require access to rich, high-quality, domain-specific data (Li, 2023); specialized human capital with deep industry knowledge (Babina et al., 2024); iterative model refinement tailored to domain-specific feedback loops (Song & Sun, 2024); and often, the co-evolution of complementary technical standards and practices (Cihon, 2019). Consequently, the degree to which a particular industry domain has integrated AI technologies—what we term **domain AI readiness**—becomes a crucial determinant of whether firms' internal AI capabilities can be successfully translated into performance gains.

The importance of domain AI readiness is observed across multiple AI-intensive industries. In financial services, the promise of AI-driven credit scoring has been tempered not only by firms' technical capabilities but also by the stringent requirements of the regulatory environment for transparency and fairness, underscoring that AI's returns depend on industry-specific governance and data standards.[1] In autonomous driving, breakthroughs in AI-based perception have proven insufficient without parallel advances in system-level engineering and sensor technologies, illustrating that progress hinges on the integration of AI with industry-specific technological infrastructures.[2] In social media content moderation, AI tools for detecting misinformation and hate speech remain reliant on extensive human oversight due to persistent accuracy and accountability concerns, revealing the critical role of domain-specific knowledge and contextual alignment (Aral, 2021). Taken together, these cases demonstrate that, unlike traditional IT adoption—where macro-level infrastructures such as bandwidth or connectivity were the primary complements—the performance outcomes of AI adoption are shaped more decisively by the degree of technological alignment and integration within the specific industry domain. Accordingly, firms embedded in domains with higher AI readiness face stronger incentives and lower barriers to developing AI

---

[1] Deloitte. State of AI in Financial Services 2023. Deloitte Insights. https://www2.deloitte.com

[2] An industry discussion article on autonomous driving introduces how the development of sensor technology complements AI-based autonomous driving algorithms. OpenCV.ai: https://www.opencv.ai/blog/ai-in-driverless-cars

capabilities. Thus, we hypothesize:

***Hypothesis 1:*** *Firms that operate in domains of higher AI readiness are more likely to increase their investment in AI capability.*

The external technological environment's level of AI readiness does not merely shape the incentives for firms to invest in AI; it also amplifies or constrains the productivity returns to those investments. This is because the effectiveness of AI hinges on the quality, accessibility, and structure of the data embedded in a given domain (Li, 2023). In domains with rich, standardized, and interoperable datasets, AI systems can more readily generate accurate predictions, identify patterns, and automate complex tasks (Wang and Wu 2024). Moreover, higher AI readiness implies lower integration costs, as firms can leverage established technical infrastructures and domain-specific metaknowledge without incurring prohibitive investments in data cleaning, translation, or system customization (Fügener et al. 2021, 2022). However, in domains where data remain fragmented, proprietary, or incompatible with AI architectures, firms face higher integration costs and greater risks of model underperformance, thereby limiting the productivity gains from AI adoption. In this sense, domain AI readiness serves not only as an enabler of adoption but also as a magnifier of returns, aligning technological opportunities with firms' internal capabilities.

For example, domain AI readiness has transformed the economics of drug discovery in the pharmaceutical industry. Traditionally, developing a new drug requires approximately US$2.6 billion and takes 12 to 15 years, with less than a 10% success rate at the clinical trial stage (DiMasi et al., 2016). The scale of the challenge—searching through an estimated $10^{60}$ possible compounds (Mullard, 2017)—renders drug discovery comparable to finding a needle in a haystack. Early attempts by pharmaceutical firms to deploy AI in drug discovery often fell short, not because of a lack of internal technical ambition, but because the external domain environment was not yet ready. Critical biomedical datasets were fragmented or inaccessible, simulation models were underdeveloped, and the integration of biological knowledge with machine learning techniques remained limited. As a result, firms experimenting with AI frequently encountered bottlenecks, with models that failed to generalize or insights that could not translate into viable compounds (Ekins, 2016; Vamathevan et al., 2019; Gold & Cook-Deegan, 2025). In other words, the absence of domain-level AI readiness—rather than the absence of firm-level initiative—constrained the

productivity of early AI adoption in drug discovery.

By contrast, the recent maturation of biomedical AI readiness has transformed the innovation landscape. The growth of large-scale biological databases, advances in computational chemistry, and cross-disciplinary research have created the conditions under which AI tools can meaningfully accelerate R&D. This shift explains why recent breakthroughs such as the AI-assisted discovery of Halicin (Stokes et al., 2020) and the acceleration of COVID-19 vaccine development (Zhang et al., 2023) succeeded where earlier efforts did not. These examples underscore a central theoretical point: AI adoption outcomes are not solely a function of firm-level capabilities, but of how those capabilities interact with the external domain. In domains of higher AI readiness, firms can more effectively leverage their AI investments, benefiting from networked data ecosystems, codified domain expertise, and established technical complementarities. Conversely, in domains with low readiness, firms encounter diminishing returns to AI capability, regardless of their internal investments. Thus, we hypothesize that:

***Hypothesis 2:*** *Firms that operate in domains of higher AI readiness achieve higher productivity from their AI capability investment.*

## 2.4 Domain–AI Integration versus Firm Strategic Pivots

Yet, the relationship between domain readiness and firm productivity raises a deeper question: to what extent can firms themselves overcome domain constraints through strategic pivots? While some firms may attempt to reorient their business models or reallocate resources to strengthen AI initiatives, their success is ultimately contingent on the readiness of the domain. There are several reasons why external domain readiness exerts a stronger influence on AI performance than internal strategic pivots. First, technological adoption is subject to strong path dependence: firms' existing routines, infrastructures, and knowledge bases constrain the extent to which they can radically reconfigure themselves to suit AI.[3] Strategic pivots often entail prohibitive costs and risks, including the disruption of established business models, organizational inertia, and the potential erosion of core competences (Dewan & Ren, 2007; Dewan & Ren, 2011; Dhyne et al., 2020). By contrast, domain-level advances—such as the accumulation of standardized datasets, the codification of domain knowledge, and the establishment of interoperable infrastructures—shift the production frontier for all firms simultaneously, lowering entry barriers and enabling a wider diffusion of AI productivity gains.

---

[3] David (1985) is the first to document the phenomenon of path dependence in technological change through his research on the history of keyboards, while Sydow et al. (2009) observed the widespread existence of organizational path dependence. Castaldi (2011) summarizes the different manifestations, sources, and consequences of technological and organizational path dependence.

Second, domain readiness reflects collective ecosystem dynamics. Many of the critical complements to AI adoption, such as shared technical standards, regulatory frameworks, and cross-firm data infrastructures, exhibit public-goods characteristics and network externalities (Cornes & Sandler, 1996). No single firm can build these conditions alone, but once established, they allow all firms to more easily realize the potential of their AI capabilities. Higher domain readiness also reduces uncertainty and risk by providing proven application templates, best practices, and benchmarks, thereby mitigating the trial-and-error costs individual firms would otherwise face in isolation.

This dynamic reflects the broader logic of technological path dependence and ecosystem complementarities. Because data infrastructures and AI-compatible technological processes evolve collectively within domains, individual firms' ability to capture value from AI hinges more on external advances in domain–AI integration than on isolated firm-level strategies. In other words, the complementarity operates more strongly between firms and their domains than within firms alone. Thus, we hypothesize that:

***Hypothesis 3:*** *The complementarity between firms' AI capabilities and performance outcomes is driven primarily by external advances in domain–AI integration rather than by firms' own strategic pivots.*

## 3. Data and Measurement

### 3.1 Data Source

We obtained more than 50 million patent application data from 2000 to 2024 in China from the Incopat database. We obtained 190 million online job posting data from 2016 to 2022 from RESSET, covering more than 500 recruitment websites and including almost all the recruitment information on the Chinese Internet. The financial data of enterprises comes from CSMAR. The data on AI policies in various provinces of China are sourced from VipLaw database of Peking University.

### 3.2 Measure Domain AI Readiness

To measure the integration level of a domain with AI, we follow the literature in technology management to use the co-occurrence of technological classes in patents, represented by IPC4 codes (Caviggioli, 2016; Lee et al., 2015). Patent IPC4 co-occurrence refers to the simultaneous appearance of two or more different IPC4 classification codes in the same patent. For instance, suppose a patent is classified under IPC4 code A61K (Preparations for medical, dental, or toilet purposes) and C07D (Heterocyclic compounds). The co-occurrence of these two IPC4 classes within the same patent suggests that the invention involves pharmaceutical preparations (A61K) based on specific chemical compounds (C07D). Since the patent IPC4 classification codes define the technical fields of patents, the co-occurrence of patent IPC4 classification codes indicates that knowledge from different technical fields is jointly used in the same patent. The more frequently two fields co-occur in patents, the greater the degree of integration between those domains.

To measure the integration level of a domain with AI, we first follow official Chinese government documents to identify IPC4 codes classified as AI-related, yielding approximately 2 million AI-related patents by 2024.[4] For each year, we then calculate how frequently each non-AI IPC4 code co-occurs with AI-related codes. This frequency serves as our baseline measure of a domain's AI readiness. To mitigate skewness of frequencies, we rank these frequencies from highest to lowest and divide them into deciles. Each decile is then mapped to a score between 1 (top decile) and 0.1 (bottom decile), which serves as our indicator of AI integration within each technological field.[5] For IPC4 codes that do not co-occur with AI-related patents in a given year, we assign a value of 0. In addition, the decile-based transformation attenuates the potential bias introduced by highly concentrated patenting activities. Specifically, even if a single firm files an exceptionally large number of AI-related patents within a domain, its disproportionate contribution to raw co-occurrence frequencies exerts only a limited effect on the decile-based ranking.

Figure 1 visualizes the co-occurrence network of IPC4 codes in AI patents. Blue nodes represent AI-

---

[4] Official documents from the National Intellectual Property Administration can be referred to: https://www.cnipa.gov.cn/attach/0/d32119ae1faa4fbf9e308824862f87dd.pdf

[5] The following results in this paper are also consistent if using the proportion, instead of frequency, of each IPC4 code's occurrence in AI patents relative to all patents for ranking.

related IPC4 codes, and green nodes represent non-AI IPC4 codes. Blue edges indicate co-occurrences among AI-related codes, while green edges capture co-occurrences between AI- and non-AI-related codes. The figure shows that, over time, AI-related technological fields have become increasingly integrated with other domains—particularly with non-AI fields—highlighting the growing breadth of AI's technological embedding.

We then calculate the domain AI readiness for each company each year, based on their patent portfolios. Specifically, the firm's domain AI readiness for the company $i$ in year $t$ is calculated as:

$$Firm\ Domain\ AI\ Readiness_{i,t} = \sum_j \left(\frac{Number\ of\ Patents\ with\ IPC\ code\ j_t}{Total\ Number\ of\ Patents_t} \times Domain\ AI\ Readiness_{j,t}\right)$$

where $j$ represents an IPC code in the company's patent portfolio. This metric measures the extent to which a company's overall technological portfolio is integrated with AI. Figure 2 shows an example of how this metric is calculated.

Figure 3 presents the average levels of domain AI readiness of firms across various industries in 2016 and 2022. Notably, industries such as manufacturing, scientific research, and social work experienced substantial increases in AI integration by 2022 compared to 2016. In contrast, sectors such as accommodation, sports, and entertainment exhibited relatively moderate growth in AI integration.

### 3.3 Measure AI Capability

Current literature has measured firm AI capability using public disclosures and talent recruitment (Lou & Wu, 2021; Babina et al., 2024; Wang & Wu, 2024). Following this practice, we construct a composite measure of a firm's AI capability from three sources: its strategic attention to AI, AI-related assets, and AI-related talent. The keywords used to identify AI technologies in this subsection can be found in Online Appendix A.

**AI strategy:** Following Tang & Xu (2025), we proxy for management's strategic attention to AI by counting the frequency of AI-related words in the Management Discussion and Analysis (MD&A) section of annual reports. Using a comprehensive vocabulary drawn from Tang & Xu (2025) and CSMAR's AI keyword list, we calculate the annual frequency of AI mentions for each firm and take the natural logarithm as the measure of AI strategy.

**AI asset:** We use keywords to screen out the fixed assets and intangible assets related to artificial intelligence from the footnotes of the financial reports of listed companies, and take the natural logarithm of their values as a measure of the AI assets of listed companies. The vocabulary list and data are compiled by the data vendor CSMAR with reference to Rammer et al. (2022) and Li et al. (2023).

**AI talent:** We assess AI talent by calculating the share of AI-related job postings in a firm's total online recruitment. To do this, we build a bilingual AI skills keyword list leveraging the work of Babina et al. (2024) and Alekseeva et al. (2021), and supplementing it with AI-related terms from the *Patent Classification System for Key Digital Technologies*.[6] Job postings are classified as AI-related if their title or description contains any keyword from the list. AI talent is then measured as the proportion of AI-related postings in total recruitment ads.

Finally, we construct an overall **AI capability** index by summing the standardized values of the three components, following Lou & Wu (2021). While some studies also use AI patents as a proxy for firm AI capability (Lou & Wu, 2021), we deliberately exclude them from our measure to avoid correlation with our domain readiness variable, which is also patent-based.

### 3.4 Measure Firm Productivity

We measure firm productivity using revenue per employee and total factor productivity (TFP), two widely-adopted metrics for the productivity of enterprises:

**Revenue per employee:** Following Tamble and Hitt (2014), we use the annual main business revenue of the enterprise divided by the total number of employees at the end of that year as a measure of labor productivity per unit.

**TFP:** We calculate firm TFP following Wooldridge (2009), which improved the estimation methods of Olley and Pakes (1996) and Levinsohn and Petrin (2003). This method proposes a one-step estimation method based on the Generalized Method of Moments (GMM), which has two advantages: (1) It overcomes the potential identification problem in the first-step estimation; (2) It can obtain robust standard errors when considering serial correlation and heteroskedasticity. Hence, we use the GMM to estimate the TFP of enterprises.

---

[6] Official documents available at:
https://www.hunan.gov.cn/zqt/zcsd/202309/29501413/files/6fc21f0aeb87465bbc3f07504a95310b.pdf

Using above methods, we construct an unbalanced firm-year panel for over 3,000 public firms in China from 2016 to 2022. The summary statistics of key variables are reported in Table 1. Among them, the average value of AI readiness is 0.477, the 1st percentile is 0, and the 99th percentile is 1, showing sufficient variation in AI readiness among enterprises. The measurements for corporate AI capability—AI strategy, AI talents, and AI assets—also show sufficient variations.

## 4. Empirical Methodology

### 4.1 Complementarity Test

In order to test the complementarities between company AI capability and domain AI readiness, we follow the previous literature and conduct two statistical tests: the demand equation that examines the correlations in inputs, and the performance test that examines the performance differences when complementary inputs are used in combination and independently (Arora & Gambardella 1990, Athey & Stern 1998, Aral & Weill 2007, Brynjolfsson & Milgrom 2013, Aral et al. 2012, Tambe 2014, Wu et al. 2020). Specifically, we first conduct the demand test by examining the correlation between a company's domain AI readiness and its AI capability using the following ordinary least square (OLS) model:

$$AI\ capability_{i,t} = \beta_0 + \boldsymbol{\beta_1} \boldsymbol{Domain\ AI\ Readiness}_{i,t} + controls + \tau_t + \varepsilon_{i,t}$$

We use the enterprise's scale (measured using total assets), capital structure and financing constraints (measured using leverage ratio), and investment incentives (measured using Tobin's q) as control variables, and use the AI strategy, AI talents, AI assets, and the synthesized AI capability as the dependent variables. Time fixed effects ($\tau_t$) are controlled for shocks that change over time. Firm fixed effects are not included in the baseline specification, as domain AI readiness varies little within firms. However, we provide tests with firm fixed effects in Online Appendix B, and the main result remains consistent.

Then we fit the following OLS model with company fixed effects ($\gamma_i$) and time dummies ($\tau_t$) to test whether AI capability and domain AI readiness form a system of complements that provides additional performance improvements in firm productivity when matched together:

$$\begin{aligned}Productivity_{i,t+1} &= \beta_0 + \boldsymbol{\beta_1} AI\ capability_{i,t} + \boldsymbol{\beta_2} Domain\ AI\ Readiness_{i,t} \\ &\quad + \boldsymbol{\beta_3} AI\ capability_{i,t} \times Domain\ AI\ Readiness_{i,t} + controls + \tau_t + \gamma_i + \varepsilon_{i,t}\end{aligned}$$

We use the revenue per employee and TFP in the next period as measures of the enterprise's productivity to examine the relationship between the current period's input and the efficiency change in the next period. The control variables are the same as above. The complementarities are supported by a positive $\beta_3$ coefficient.

## 4.2 Identification Strategy

Offering conclusive causal evidence proves challenging due to the endogeneity of organizational practices in observational data. The complementarity approaches used in this study are naturally robust to some types of endogeneity and reverse causality problems because they are about matching two or more investment decisions rather than about the effectiveness of the decisions themselves. Any biases that affect the complementarity term must be present only at the confluence of both factors, and not when factors are present individually (Tambe et al. 2012). In addition, the two tests for complementarities are subject to different sources of bias, so consistent results provide greater confidence that the results are not primarily due to endogenous issues (Wu et al. 2020).

In addition to the complementarity tests, we also employ supplementary identification strategies for the variables Domain AI Readiness and AI Capability. Specifically, we decompose Domain AI Readiness into an internal part that is subject to the firm's strategic pivots and an external part related to the evolving technological compatibility, which is exogenous to the firm's strategic decisions. Regarding AI Capability, we utilize AI policy changes in China, which provide exogenous variations in company AI investments.

### *4.2.1 Decomposing Domain AI Readiness*

As shown in Figure 2, the variation in domain AI readiness within firms arises from two distinct sources. The first source is the changing frequency with which IPC4 codes co-occur with AI-related patents in a given year, which is calculated using all patents in the market and is exogenous to any individual firm's strategic choices. The second source stems from shifts in the IPC4 codes represented in a firm's own patents, which may reflect endogenous strategic pivots. To disentangle these mechanisms and mitigate potential endogeneity, we decompose domain AI readiness into two components. The **Internal Strategic Pivot** component is constructed by pooling all AI-related patents from 2016 to 2022 to calculate the readiness level of each IPC4 code, thereby holding the external technological environment constant over time. The **External Technological Evolvement** component is constructed by pooling all patents of the focal firm from 2016 to 2022 to calculate its patent portfolio, thereby holding the firm's internal portfolio constant over time. By construction, the external component is immune to the time-varying patenting activity of any single company and is thereby exogenous to the company's strategic decisions. Thus, we treat this variable as exogenous in the regression model. Figure 4 presents the decomposition method using a visualized example.

### 4.2.2 Regional AI policy as instrumental variable

Firms may endogenously decide how much AI investment to make. Drawing on the approach of Wang and Wu (2024), we use the AI-promoting policies in the region where the firm is located as the instrumental variable for the firm's AI capabilities. As Wang and Wu (2024) pointed out, many local governments in China have offered incentives, such as cash subsidies and tax abatements, to encourage local businesses to adopt AI over the past decade. These incentives are often unexpected, making them an exogenous source of variation useful for studying company performance. Thus, we create an instrumental variable for a firm's AI investment using the cumulative number of active AI policies in the province where the company is located, measured at the end of each year. We identified a total of 1,343 policies containing AI-related keywords from VipLaw database of Peking University from 2016 to 2022. We lag the instrumental variable by one year since policies may be promulgated at any time during the year, and it usually takes some time for them to have an impact on enterprises.

This instrumental variable varies at the region-year level. The geographic distribution and time variance provide sufficient power for the first-stage analysis of the two-stage least squares (2SLS) model. To support the validity of our instrument, we examine whether the instrumental variables predict future productivity levels in companies that have not invested in AI, following Martin & Yurukoglu (2017) and Wang et al. (2024). The logic is as follows: if the instrument is valid, then it should be correlated only to the focal company's productivity through its effect on the focal company's AI capability. For companies without AI investments, we should not observe a significant correlation between the instrument and the outcome variable. To test this, we regress the dependent variable directly on the instrumental variables, using only data from observations that did not involve AI talent hiring or AI assets. The results shown in Online Appendix D support the validity of the instrumental variable.

## 5. Results and Interpretation

We first show the demand test results in Table 2. After adding controls, the relationships between the degree of AI readiness and the four different measures of AI capabilities, including AI strategy, AI talents, AI asset, and the integrated measure of AI capability, are all statistically significant at the 1% level. This suggests that companies with higher domain AI readiness also pay more attention to and invest more in AI capabilities, supporting H1.

We then show the performance test results in Table 3. Columns 1 and 4 in Table 3 present the baseline regression results, and the coefficient of the interaction term between **AI readiness** and **AI capability** is positive. This indicates that when the degree of domain AI readiness in an enterprise is high, the efficiency improvement brought about by investing in AI is higher than when the degree of domain AI readiness in the enterprise is low, supporting H2. For firms with the same level of AI capability, those with a domain AI readiness of 1 have labor productivity and total factor productivity that are 3% and 2% standard deviation higher, respectively, than those with a domain AI readiness of 0. Meanwhile, the regression coefficient of **AI capability** alone is negative when controlling for the interaction term, suggesting that investing in AI capability in domains of low readiness can negatively impact the innovation performance in companies; however, if increasing AI capability in domains of high readiness, the overall effect is positive. Overall, the results suggest that when an enterprise considers investing in AI capabilities, it should pay attention to the degree of integration between its current technological field and AI.

Next, we decompose the **AI readiness** into internal and external parts. Columns 2 and 5 of Table 3 report the regression results focusing solely on the variation in the external technological evolvement. Notably, the coefficient of the interaction term in these columns is approximately threefold economically compared to the baseline estimate, accompanied by a significant enhancement in statistical significance. This amplification suggests that the complementary effects are mainly driven by the external technological evolvement, supporting H3. In contrast, Columns 3 and 6 focus on the variation of the internal strategic pivots, where the interaction terms are no longer statistically significant, indicating that shifts in the patenting domains alone do not contribute meaningfully to the complementary effect. Such results also alleviate potential endogeneity concerns, such as better companies strategically shifting their domains to maximize the benefits of their AI investment. We also conduct further robustness tests, including running the regression model using an alternative measurement of AI readiness and excluding enterprises that have applied for AI patents. The results presented in Online Appendix C are consistent with the main findings.

Table 4 reports the results of the 2SLS estimation using instrumental variables.[7] Column 1 shows the first-stage results. The regression coefficient of the AI policy is significantly positive at the 1% level, which means that after the local government issues AI-related policies, the AI capabilities of local enterprises have significantly improved. Column 2 and Column 3 present the results of the second-stage estimation. The regression coefficients of the interaction term remain significantly positive, consistent with the above results. Note that the coefficients in the 2SLS regressions have increased considerably by about 10 to 20 times in comparison to the OLS results, suggesting that the magnitudes of the effects we observe are in reality larger than what is suggested by the OLS results.

Collectively, these findings provide robust evidence that the complementary relationship between domain AI readiness and AI capability primarily stems from external advancements in domain-AI integration, as reflected by changes in IPC4 co-occurrences, rather than from enterprises' strategic shifts in technological fields. The results thus offer prescriptive implications for corporate strategy: enterprises are advised to adopt a strategic waiting approach, deferring AI investments until the technology matures sufficiently to align with their existing technological capabilities. Proactive realignment of technological fields merely to accommodate AI technologies is shown to be ineffective in enhancing the complementary benefits of AI adoption.

## 6. Mechanism Tests

In this section, we first examine the source of internal productivity gain of firms after investing in AI

---

[7] We employ various statistical test methods to verify the effectiveness of the instrumental variables: Firstly, we use the Kleibergen-Paap rk LM statistic for the identification test. The results are all significant at the 1% level, rejecting the insufficient identification hypothesis. Secondly, we use the Cragg-Donald Wald F statistic for the weak instrumental variable test. The results are all greater than the critical value at the 15% level, rejecting the weak instrumental variable hypothesis. We also test the exclusivity of instrumental variables following Martin & Yurukoglu (2017) and Wang et al. (2024), and the results are presented in Online Appendix D.

capability. Then we examine the source of the external evolution of domain-AI integration level and offer some stylized evidence.

**6.1 Productivity Gain through Product Innovations**

Product innovation is crucial for enterprises to improve productivity as it enables them to identify and meet unmet market needs, driving revenue growth while reducing the risk of resource waste on inefficient products (Krishnan & Loch, 2005; Bhaskaran et al., 2021). Literature has documented that product innovation has been the main driver of performance improvement after companies adopt AI (Babina et al. 2024). When combined with AI, product innovation becomes even more powerful as AI can automate large-scale data analysis and uncover hidden patterns to guide new product development (Verganti et al. 2020, Berente et al. 2021; Wang & Wu, 2024). For example, Insilico Medicine used AI to analyze data, identify DDR1 kinase inhibitors, and discovered six new compounds for fibrosis treatment in 46 days, demonstrating AI's efficiency in accelerating drug-target identification and product innovation (Lou & Wu 2021). However, the ability of AI to empower product innovation is also related to the attributes of the product. The capability of AI-assisted new drug discovery may not be simply transferred to the discovery of new metal materials (Krieger et al. 2022),[8] which is related to whether the characteristics of the drug and metal material domains can be well integrated with AI technologies. Therefore, we expect that whether AI can promote product innovation is also related to the AI readiness in the enterprise's own technical field; that is, when the AI readiness of the enterprise is high, the enterprise's AI capabilities can promote product innovation.

We measure the number of new products using the number of product trademarks that a company applies for each year (Babina et al., 2024; Gao & Hitt, 2012). This measure is useful because it is a common practice for companies in China to protect intellectual property before launching new products (Wang & Wu, 2024). We replace the dependent variable in Equation (1) with the number of trademarks and use Poisson regression to adapt to the nature of the count variable of trademark applications.

The regression results are shown in Table 5. Overall, we find that the regression coefficient of the interaction term is significant. Moreover, the coefficient of the interaction term between **AI readiness (external technological evolvement)** and **AI capability** in Column 2 is more significant than that between **AI readiness** and **AI capability** in Column 1, consistent with previous results. Additionally, we find that the regression coefficient of **AI capability** alone is positive but not statistically significant, indicating that the ability of AI to empower firms' product innovation is indeed related to the firms' AI readiness. When the domain AI readiness of a firm is very low, the promoting effect of AI on the firm's product innovation may be minimal.

---

[8] Similar discussions can be found in the following blog and a series of tweets by Robert Palgrave: https://thebsdetector.substack.com/p/ai-materials-and-fraud-oh-my

## 6.2 External Source of Domain-AI Integration

### 6.2.1 Temporal Dynamics of IPC4 Deciles

*Domain AI readiness* is composed of two interrelated elements. One is the technical field of the patents applied by the enterprise, which is closely tied to its core business. The other is the quantile of IPC4 co-occurrence frequencies. In this subsection, we briefly discuss how the quantiles of IPC4 have changed during the sample period.

We start with the growth in the distribution of different IPC classes. Figure 5 shows the growth of the co-occurrence counts of each IPC class in AI patents, with the values in 2016 standardized to 100. The results show that Class A (Human Necessities) experienced the largest growth in AI patents, increasing to over 10 times the original amount in seven years. Next is Class G (Physics), which grew to approximately 9 times the original amount, and Class H (Electricity) had the lowest growth rate but still increased to twice the original amount. Most of the IPC class numbers assigned to artificial intelligence fall under Section G (Physics) and Section H (Electricity). This indicates that AI has indeed been widely integrated with other technical fields, rather than being confined solely within a closely related class.

Figure 6 depicts the dynamic shifts in the decile distribution of IPC4 codes spanning from 2016 to 2022. The left-hand side represents the decile distribution of IPC4 codes in 2016, and the right-hand side corresponds to that in 2022. The multicolored flow lines serve as a visual indicator of the transitions in decile positions of IPC4 codes over these years. Notably, while the highest (Decile 10) and lowest (Decile 1) deciles exhibit relatively minor fluctuations, the intermediate deciles (Decile 2 - 9) have undergone more substantial rearrangements. This pattern reveals that the distribution characteristics of IPC4 codes within patents are not static but rather experience a continuous evolutionary process. B23Q, which pertains to machine tool parts, components, accessories, general-purpose machine tools characterized by specific structural features, and combined machine tools in the IPC classification, stands out as the IPC4 category with the most rapid growth, advancing from Decile 2 to Decile 10 over this period.

### *6.2.2 Rapid Growth of Academic Papers and IPC4 Co-Occurrences in Biological Fields*

Since the decile ranking of IPC4 codes exhibits substantial variation over time, we next investigate the factors driving these changes. A natural candidate is the academic advancement of AI technologies. Breakthroughs such as BERT in natural language processing and ResNet in image recognition have spurred corresponding applications,[9] suggesting that increases in the co-occurrence of specific IPC4 codes with AI patents may be closely linked to advances in related research domains. While the rapid progress of AI is evident, the domains where it is most extensively applied remain less clearly delineated. For example, image recognition techniques can be embedded in medical imaging for cancer detection as well as in surveillance systems for facial recognition. To systematically map AI's academic frontiers, we analyze over 200 million papers in the OpenAlex database. Between 2016 and 2022, AI-related publications in medicine and biology exhibited the fastest growth.[10] Building on this finding, we expect that if academic and patenting activities are correlated, then IPC4 categories related to medicine and biology should display a higher growth rate of AI patents relative to other domains.[11]

We use the following specifications to identify the differences in the growth rates of AI patents corresponding to biological and medical-related IPC4 and the control group:

$$\#Patents_{j,t} = \beta_0 + \boldsymbol{\beta_1} Bio-related\ IPC4_j + \boldsymbol{\beta_2} Year\ trend_t \\ + \boldsymbol{\beta_3} Bio-related\ IPC4_j \times Year\ trend_t + \varepsilon_{i,t}$$

Among them, $\#Patents_{j,t}$ represents the co-occurrence count of IPC4 j in year t. $Bio-$

---

[9] As of June 17, 2025, the Google citation counts for the BERT (Devlin et al., 2019) and ResNet (He et al., 2016) papers have reached 133,000 and 273,000 respectively, demonstrating their enormous impact on related academic fields.

[10] The thematic analysis process for AI papers in OpenAlex is described in Online Appendix F.

[11] We attempted to map IPC4 classifications to paper subjects, but found significant discrepancies between academic subject divisions and patent IPC classifications. Most IPC4 categories in patents belong to engineering fields such as textiles, papermaking, machinery, and civil engineering, whereas paper subjects lean more toward scientific disciplines like biology, physics, mathematics, and chemistry.

$related\ IPC4_j$ is a dummy variable, taking the value of 1 when the IPC4 is G16H or G16B, and 0 when it is other IPC4s in Section G. G16B refers to bioinformatics and G16H refers to healthcare informatics. These two IPC4 classes are the bio- and medical-related IPC4 classes within Section G, with other IPC4 classes in Section G serving as the control group. $Year\ trend_t$ represents the numerical value of the year. Under this regression specification, β₂ captures the linear growth rate of the control group (i.e., the co-occurrence counts of other IPC4s in Section G in AI patents), while β₃ captures the additional increment in the linear growth rate of G16H and G16B relative to the control group. Since the dependent variable is a count variable, we use Poisson regression for estimation.

Table 6 presents the estimation results. The results in Column 1 show that the co-occurrence counts of G16B and G16H grew significantly faster than other IPC4 classes in Section G from 2016 to 2022, with an increase as high as 59%. In Column 2, we estimate a specification that includes IPC4 fixed effects and year fixed effects, and the regression coefficient of the interaction term is also significant.

Through the analysis of the growth rates of biological and medical-related IPC4 categories, we provide some stylized facts indicating that the ranking of IPC4 classifications is significantly influenced by academic development. The next subsection, which explores the spillover effects among patent-filing institutions, further substantiates this argument with additional evidence.

### 6.2.3 Academic Spillovers in AI Patenting

Academic research has long been recognized as an important driver of local innovation through knowledge spillovers (Jaffe, 1989; Acs et al., 1994; Audretsch & Feldman, 1996). In China, university–industry collaboration is pervasive (Hsu et al., 2024), raising the question of whether similar spillovers occur in the integration of AI technologies into specific IPC4 domains. Patent applicants in China fall into four categories: individuals, enterprises, research institutions, and government agencies. We classify AI patents by applicant type and construct IPC4–year panel data.

Panel A of Table 7 shows that enterprises account for 70% of AI patent applications, followed by research institutions (22%), individuals (6%), and government agencies (1%). Thus, enterprises dominate AI patenting, though research institutions remain a significant contributor. Panel B further examines spillover dynamics. Column 1 regresses enterprise AI patent applications on research institution applications in the preceding three years and finds that increases in research institution patents containing a given IPC4 predict subsequent enterprise patents in the same IPC4. This pattern suggests a diffusion effect from research institutions to enterprises. Column 2 conducts a placebo test using individual patents as the independent variable and finds no similar effect, ruling out the possibility that results are driven by the general trend of AI integration across all IPC4 codes.

Taken together, these results indicate that while enterprises are the main source of AI patenting, the academic sector indirectly shapes the trajectory of IPC4 rankings by seeding technologies that later diffuse into industry.

## 7. Conclusion and Discussion

This study is motivated by the growing paradox in the digital economy: while firms increasingly invest in artificial intelligence (AI) to drive innovation and operational efficiency, empirical evidence often shows that such investments do not consistently yield the expected performance gains. Existing literature has primarily focused on internal organizational complements to AI adoption but has overlooked how external technological environments may condition the effectiveness of AI initiatives. Recognizing this gap, we seek to understand how domain-specific technological compatibility—captured through the concept of domain AI readiness—interacts with firm-level AI capabilities to influence firm performance.

Leveraging a comprehensive panel dataset of Chinese listed firms, patent records, online job postings, and regional AI policies, we find that firms' AI capabilities and domain AI readiness are highly complementary. Firms achieve significantly greater innovation outputs and productivity gains from their AI investments when operating in domains with higher AI readiness. Furthermore, by disentangling the sources of variation in AI readiness, we demonstrate that this complementarity is primarily driven by advances in external technological evolution, rather than by firms' endogenous shifts in domains. Instrumental variable analyses based on regional AI policy shocks confirm the robustness of these findings against endogeneity concerns.

Our findings highlight a fundamental aspect of AI deployment that differentiates it from previous generations of GPTs, such as traditional IT systems. Unlike IT, whose value realization was heavily influenced by macro-level infrastructure conditions, AI's value creation is deeply contingent on micro-level, domain-specific technological ecosystems. Firms with strong internal AI capabilities cannot realize the full benefits of AI unless they operate within domains where AI technologies are sufficiently embedded, mature, and integrated with industry-specific practices and standards. This underscores the critical importance of external technological alignment as a complement to internal digital transformation efforts.

Our study opens up several promising avenues for future research. First, future studies could explore how AI readiness evolves over time and affects dynamic AI capability development within firms. Second, research could examine cross-national differences in AI readiness and assess how institutional environments interact with technological ecosystems to influence AI value realization. Third, future work could refine measures of domain readiness by incorporating real-time data streams, such as digital platform usage, collaborative project data, or cross-industry partnerships. Fourth, scholars could investigate whether certain types of firms (e.g., startups vs. incumbents, digital natives vs. traditional firms) benefit differentially from domain-level technological readiness.

This study contributes to the IS literature in several ways. First, it advances the theoretical understanding of complementarity by shifting attention from purely internal organizational complements to the role of external technological environments. Second, it refines the conceptualization of AI as a GPT by highlighting its unique reliance on domain-specific technological ecosystems. Third, it introduces and operationalizes the novel concept of domain AI readiness, providing a measurable, dynamic construct for future empirical work. Overall, the study calls for a more context-sensitive, multi-level perspective in examining the business value of AI and other emerging digital technologies.

From a managerial perspective, our findings suggest that firms should not view AI investment solely as an internal capability-building exercise. Instead, firms must carefully assess the technological maturity and AI readiness of their industry domains when formulating AI strategies. Managers should prioritize AI deployment in domains where external technological conditions are favorable and consider "strategic waiting" approaches in domains where AI readiness remains low.

For policymakers, our results emphasize the importance of fostering domain-specific technological ecosystems to support AI diffusion and value creation. Beyond general support for AI research and development, targeted policies that enhance data accessibility, promote industry-specific standards, and facilitate talent development in key domains are likely to yield greater economic benefits. Policymakers should recognize that successful AI adoption requires not only firm-level investments but also a supportive, domain-specific external technological infrastructure.

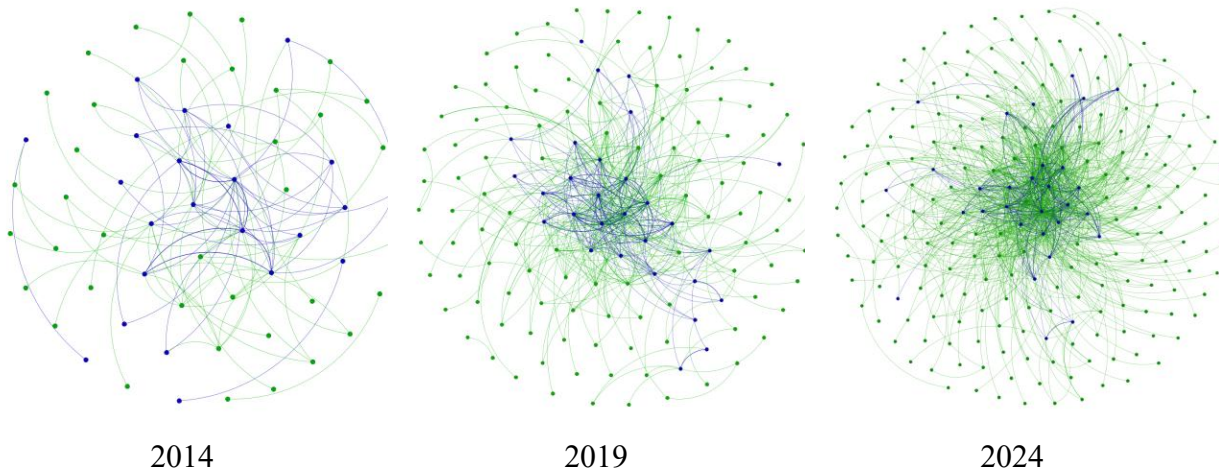

2014　　　　　　　　　　　2019　　　　　　　　　　　2024

**Figure 1. The IPC4 classification code co-occurrence network diagram in AI patents**

*Note*: This figure illustrates the network relationship between AI IPC4 and non-AI IPC4. Here, blue nodes represent AI IPC4, while green nodes denote non-AI IPC4. Blue edges indicate co-occurrence within AI IPC4, meaning that both connected nodes belong to AI IPC4. Green edges, conversely, indicate the co-occurrence of non-AI IPC4 in AI patents, where at least one of the connected nodes is non-AI IPC4. For simplicity of the network diagram, we have only included the IPC4 codes that appear more than 100 times in AI patents each year.

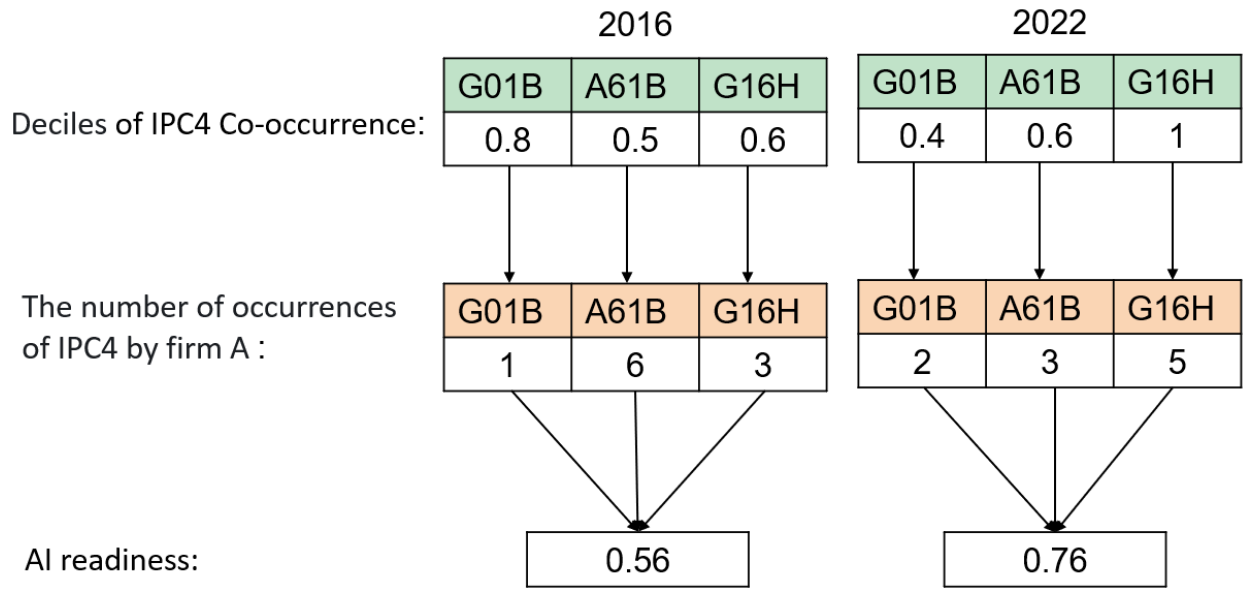

**Figure 2. Schematic diagram of the calculation process of AI readiness**

*Note*: This figure takes a fictional Company A as an example to demonstrate the calculation process of the AI readiness variable. Taking 2016 as an example, the enterprise applied for 1 patent in the IPC4 field of G01B, 6 patents in the IPC4 field of A61B, and 3 patents in the G16H field. The decile values of co-occurrence in AI patents for these three IPC4 fields are 0.8, 0.5, and 0.6 respectively. Then the AI readiness index of the company is $\frac{0.8\times1+0.5\times6+0.6\times3}{1+6+3}=0.56$.

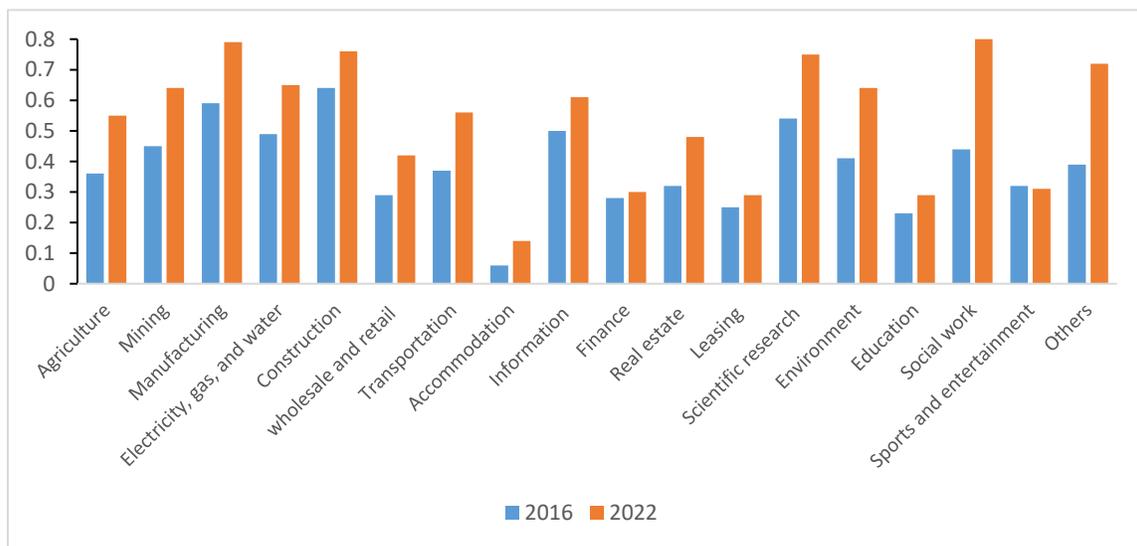

**Figure 3 Average AI readiness across industries (2016 and 2022)**

*Note*: The figure presents the average AI readiness of various industries in 2016 and 2022, calculated using a simple arithmetic mean. The industry classification follows the Industry Classification Standards issued by the China Securities Regulatory Commission (CSRC) in 2012.

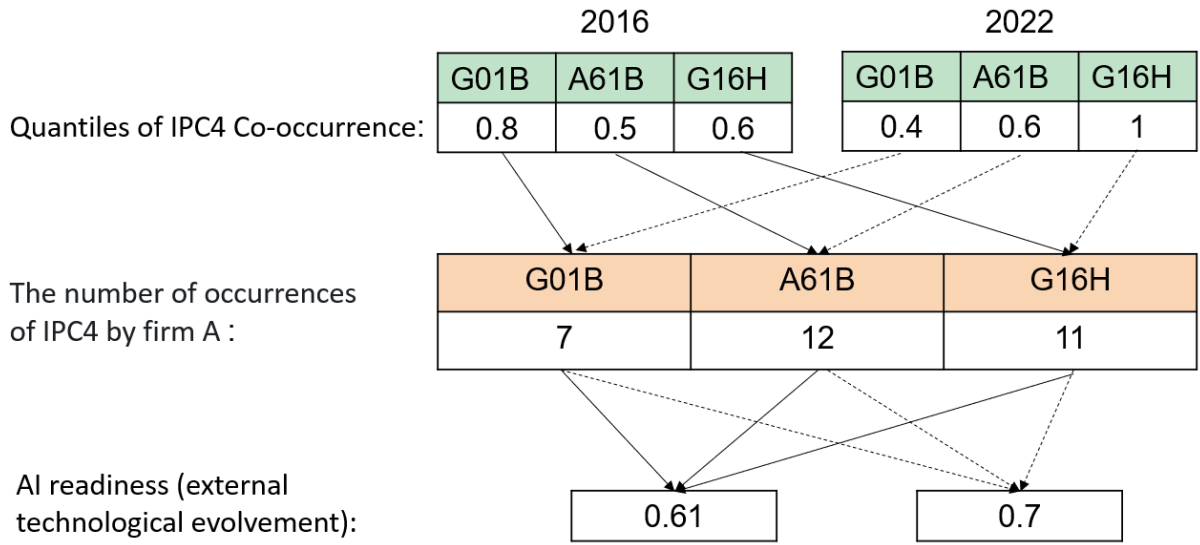

(A) Measure the change from the evolution of AI technology

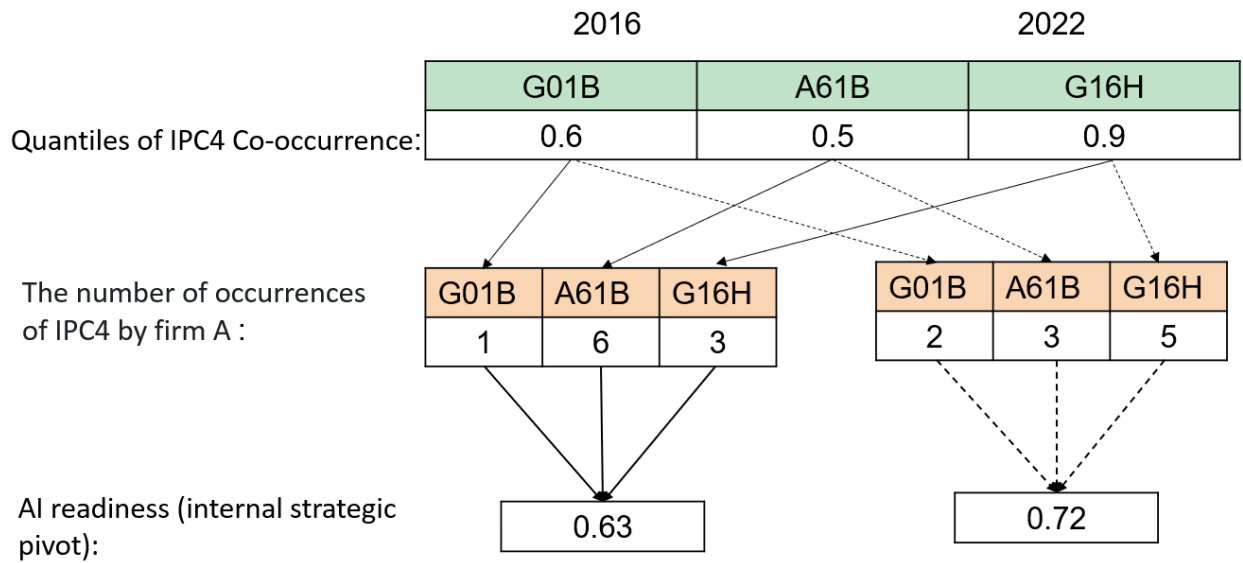

(B) Measure the change from the technological evolution of enterprises

Figure 4 Schematic diagram of the decomposition of AI readiness

*Note*: When calculating AI readiness (external technological evolvement) in Figure A, we take the IPC4 fields of all patents applied by Company A from 2016 to 2022 as a pool, allowing only the decile values of co-occurrence times of IPC4 in AI patents to vary, with the calculation method similar to that of AI readiness. The solid line represents the calculation process of AI readiness (external technological evolvement) in 2016, and the dashed line represents that in 2022. When calculating AI readiness (internal strategic pivot) in Figure B, we take the co-occurrence times of IPC4 in AI patents from 2016 to 2022 as a whole to calculate the decile, allowing only the occurrence times of IPC4 in the company's annual patent applications to vary. Similarly, the solid line denotes the calculation process of AI readiness (internal strategic pivot) in 2016, and the dashed line denotes that in 2022.

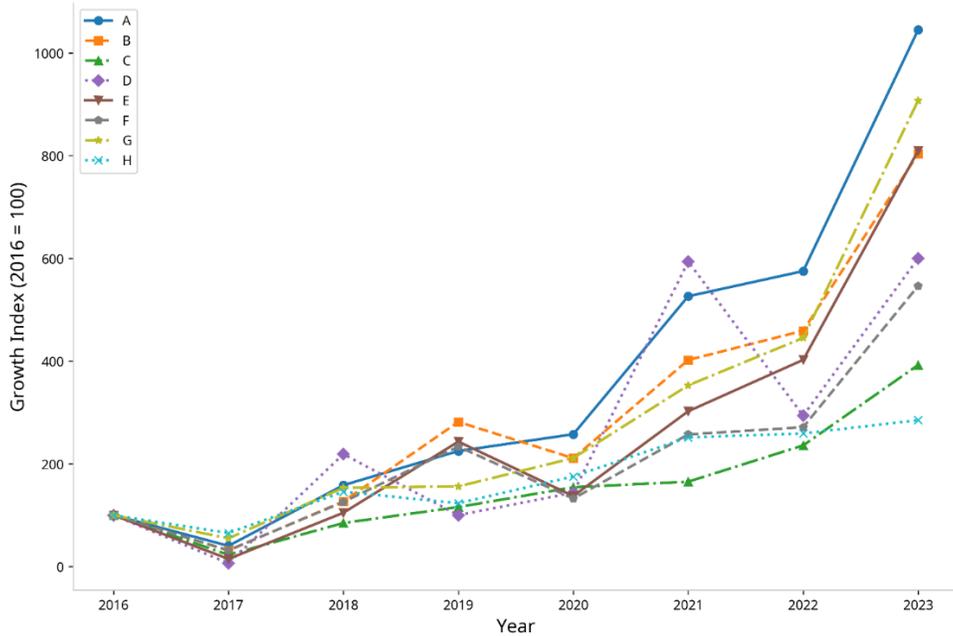

**Figure 5 The growth trends of each section of IPC in AI patents**

*Note*: The figure shows the growth of co-occurrence counts of each IPC section in AI patents from 2016 to 2023, standardized by setting the co-occurrence counts of each section in AI patents in 2016 as 100. Section A is for human necessities, B for performing operations and transporting, C for chemistry and metallurgy, D for textiles and paper, E for fixed constructions, F for mechanical engineering, lighting, heating, weapons and blasting, G for physics, and H for electricity.

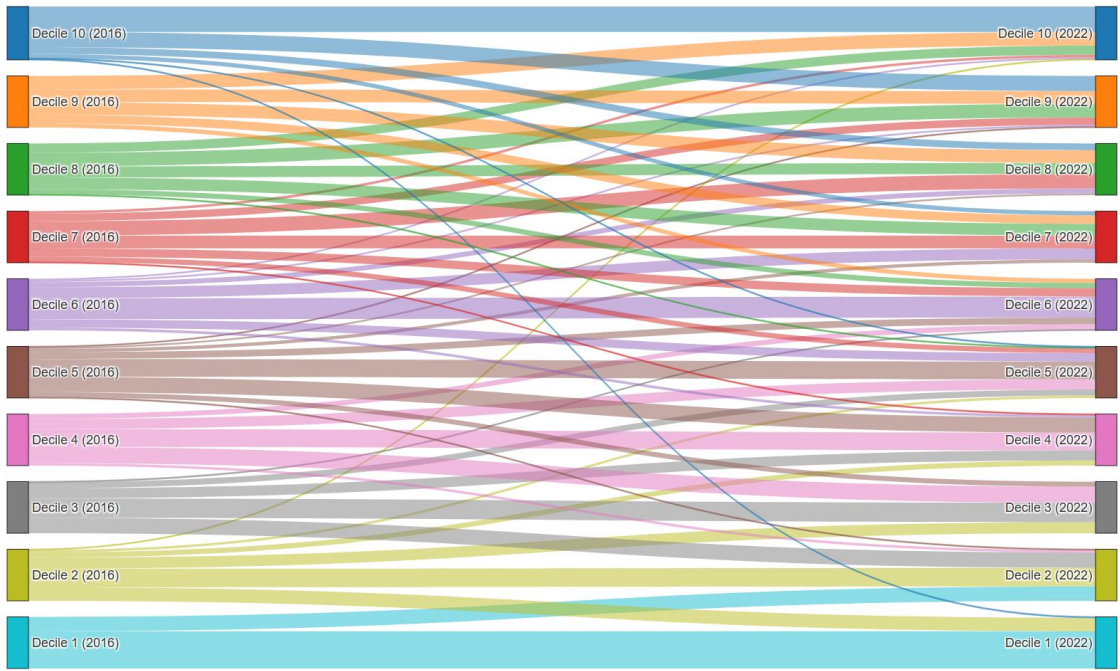

**Figure 6. Evolution Diagram of Decile Distribution of IPC4 Codes**

*Note*: The Sankey diagram shows the changes in the deciles of IPC4 from 2016 to 2022.

**Table 1 Summary statistics**

| Variables | Obs | Mean | Std.Dev | P1 | P99 |
|---|---|---|---|---|---|
| AI strategy | 20,415 | 0.087 | 0.212 | 0 | 0.954 |
| AI talents | 20,415 | 0.312 | 0.783 | 0 | 3.555 |
| AI asset | 20,412 | 19.127 | 3.483 | 0 | 24.439 |
| AI capability | 20,412 | 0.166 | 0.626 | -1.616 | 1.988 |
| AI readiness | 20,415 | 0.477 | 0.278 | 0 | 1 |
| AI readiness (external technological evolvement) | 20,415 | 0.479 | 0.286 | 0 | 1 |
| AI readiness (internal strategic pivot) | 20,415 | 0.466 | 0.283 | 0 | 1 |
| Lagged AI Policy | 16,562 | 2.221 | 1.274 | 0 | 5 |
| Revenue per employee | 20,411 | 12.981 | 0.835 | 12.236 | 16.469 |
| TFP | 20,031 | 5.794 | 0.834 | 4.011 | 8.096 |
| Trademarks | 20,415 | 12.725 | 31.907 | 0 | 186 |
| Log (Total assets) | 20,415 | 22.428 | 1.333 | 20.060 | 26.423 |
| Leverage | 20,415 | 0.429 | 0.199 | 0.065 | 0.905 |
| Tobin'q | 20,415 | 1.982 | 1.317 | 0.843 | 8.563 |

*Note*: P1 represents the 1st percentile, and P99 represents the 99th percentile.

**Table 2 AI readiness and AI capability**

|  | (1) AI strategy | (2) AI talents | (3) AI asset | (4) AI capability |
|---|---|---|---|---|
| AI readiness | 0.0835*** | 0.4991*** | 0.9088*** | 0.3375*** |
|  | (0.0080) | (0.0276) | (0.1482) | (0.0290) |
| Log (Total assets) | 0.0032 | 0.0485*** | 0.8111*** | 0.0271*** |
|  | (0.0024) | (0.0114) | (0.0660) | (0.0102) |
| Leverage | -0.0242* | -0.1067** | -0.5225* | -0.0627 |
|  | (0.0130) | (0.0464) | (0.3135) | (0.0500) |
| Tobin'q | 0.0052*** | 0.0574*** | -0.0598 | 0.0411*** |
|  | (0.0019) | (0.0086) | (0.0441) | (0.0072) |
| Year FE | Yes | Yes | Yes | Yes |
| # of Obs | 20,415 | 20,415 | 20,412 | 20,412 |
| $R^2$ | 0.0412 | 0.0978 | 0.1091 | 0.0403 |

*Note*: Standard errors are clustered at the firm level and reported in parentheses. * p<0.10, ** p<0.05, *** p<0.01.

**Table 3 Tests for complementarity between AI capability and AI readiness**

|  | (1) | (2) | (3) | (4) | (5) | (6) |
|---|---|---|---|---|---|---|
|  | Lead. Revenue per employee | | | Lead. TFP | | |
| AI readiness | 0.0028 | | | 0.0029 | | |
|  | (0.0077) | | | (0.0074) | | |
| AI readiness (external technological evolvement) | | 0.0153 | | | 0.0111 | |
|  | | (0.0170) | | | (0.0161) | |
| AI readiness (internal strategic pivot) | | | -0.0001 | | | -0.0000 |
|  | | | (0.0001) | | | (0.0001) |
| AI capability | -0.0297** | -0.0774*** | -0.0345* | -0.0299** | -0.0572*** | -0.0224 |
|  | (0.0128) | (0.0244) | (0.0207) | (0.0118) | (0.0199) | (0.0181) |
| AI readiness×AI capability | **0.0272*** | | | **0.0157*** | | |
|  | **(0.0101)** | | | **(0.0090)** | | |
| AI readiness (external technological evolvement) ×AI capability | | **0.0822*** | | | **0.0480**** | |
|  | | **(0.0274)** | | | **(0.0242)** | |
| AI readiness (internal strategic pivot) ×AI capability | | | **0.0202** | | | **-0.0031** |
|  | | | **(0.0268)** | | | **(0.0240)** |
| Controls | Yes | Yes | Yes | Yes | Yes | Yes |
| Firm FE | Yes | Yes | Yes | Yes | Yes | Yes |
| Year FE | Yes | Yes | Yes | Yes | Yes | Yes |
| # of Obs | 16,417 | 16,417 | 16,417 | 16,417 | 16,099 | 16,099 |
| $R^2$ | 0.8664 | 0.8665 | 0.8663 | 0.8664 | 0.8793 | 0.8793 |

*Note*: Standard errors are clustered at the firm level and reported in parentheses. * $p<0.10$, ** $p<0.05$, *** $p<0.01$.

**Table 4 IV approach**

| Dependent variable: | (1) AI capability | (2) Lead. Revenue per employee | (3) Lead. TFP |
|---|---|---|---|
| AI readiness (external technological evolvement) |  | 0.5584 (0.3686) | 0.5473 (0.3687) |
| AI capability (estimated) |  | -1.1668** (0.5891) | -1.1839** (0.5895) |
| AI readiness (external technological evolvement)×AI capability (estimated) |  | **1.0436*** (0.3659)** | **1.1070*** (0.3559)** |
| Lag. AI policy | 0.0206** (0.0088) |  |  |
| Controls | Yes | Yes | Yes |
| Firm FE | Yes | Yes | Yes |
| Year FE | Yes | Yes | Yes |
| F-statistics | 27.23 | 24.49 | 42.63 |
| Kleibergen-Paap rk LM statistic |  | 10.086 | 9.729 |
| Cragg-Donald Wald F statistic |  | 8.371 | 8.147 |
| # of Obs | 14,212 | 14,155 | 13,884 |
| $R^2$ | 0.7694 | - | - |

*Note*: Standard errors are clustered at the firm level and reported in parentheses. * $p<0.10$, ** $p<0.05$, *** $p<0.01$.

**Table 5 Test for corporate product innovation**

|  | (1) | (2) |
|---|---|---|
| Dependent Variable: | Lead. #Trademarks | |
| AI capability | 0.0507 | 0.0110 |
|  | (0.0915) | (0.0908) |
| AI readiness | -0.1154 |  |
|  | (0.0845) |  |
| AI readiness×AI capability | **0.1930*** |  |
|  | **(0.1077)** |  |
| AI readiness (external technological evolvement) |  | -0.0838 |
|  |  | (0.0878) |
| AI readiness (external technological evolvement)×AI capability |  | **0.2361**** |
|  |  | **(0.1083)** |
| Controls | Yes | Yes |
| Firm FE | Yes | Yes |
| Year FE | Yes | Yes |
| # of Obs | 10,815 | 10,815 |

*Note*: Standard errors are clustered at the firm level and reported in parentheses. * p<0.10, ** p<0.05, *** p<0.01. The model used in the regression is Poisson regression.

**Table 6 The growth of biology- and medicine-related IPC4**

|  | (1) | (2) |
|---|---|---|
| Dependent Variable: | #Patents | |
| Bio-related IPC4 | -323.219** |  |
|  | (159.525) |  |
| Year trend | 0.2701*** |  |
|  | (0.0567) |  |
| Bio-related IPC4×Year trend | **0.1594**** | **0.1798**** |
|  | **(0.0789)** | **(0.0362)** |
| IPC4 FE | No | Yes |
| Year FE | No | Yes |
| # of Obs | 850 | 800 |

*Note*: Standard errors are clustered at the firm level and reported in parentheses. * p<0.10, ** p<0.05, *** p<0.01. The model used in the regression is Poisson regression.

## Table 7 The Diffusion of AI Innovation from Research Institution to Enterprises

Panel A: AI patent applications by applicant type

| Applicant Type | Average patent count per IPC4 | Proportion |
|---|---:|---:|
| Individuals | 12.78 | 6.08% |
| Enterprises | 148.36 | 70.65% |
| Research institutions | 46.07 | 21.94% |
| Government | 2.8 | 1.33% |
| Total | 210 | 100% |

Panel B: The correlation between the number of AI patents applied for by different AI patent applicants

|  | (1) | (2) |
|---|:---:|:---:|
| Dependent variable: | #Patents by enterprises | |
| Lag. #Patents by research institutions | 0.0006*** | |
|  | (0.0002) | |
| Lag2. #Patents by research institutions | 0.0006** | |
|  | (0.0003) | |
| Lag3. #Patents by research institutions | -0.0001 | |
|  | (0.0001) | |
| Lag. #Patents by individuals |  | 0.0001 |
|  |  | (0.0001) |
| Lag2. #Patents by individuals |  | 0.0001 |
|  |  | 0.0001 |
| Lag3. #Patents by individuals |  | -0.0005*** |
|  |  | (0.0001) |
| IPC4 FE | Yes | Yes |
| Year FE | Yes | Yes |
| # of Obs | 13,068 | 13.068 |
| Pseudo $R^2$ | 0.9591 | 0.9538 |

*Note*: Panel A shows the average annual IPC4 co-occurrence frequency in AI patent applications filed by different types of applicants. Panel B, conversely, conducts a regression analysis where the dependent variable is the IPC4 co-occurrence frequency in enterprises' AI patent applications. The independent variables included are the lagged terms of IPC4 co-occurrence frequencies from both research institutions' and individual applicants' AI patent applications. Standard errors are clustered at the firm level and reported in parentheses. * $p<0.10$, ** $p<0.05$, *** $p<0.01$. The model used in the regression is Poisson regression.

# Online Appendix A. Word list for AI strategy, AI assets and AI talents

Table A1 presents the word list we use to identify AI strategies in the MD&A section of annual reports, AI talents in job postings, and AI assets in the asset breakdown items of annual reports.

Table A1 AI keyword glossary

| Variable | AI keyword |
| --- | --- |
| AI strategy | Artificial Intelligence (AI), Business Intelligence (BI), Image Understanding, Investment Decision-making Assistance System, Intelligent Data Analysis, Intelligent Robot, Machine Learning, Deep Learning, Semantic Search, Biometric Technology, Face Recognition, Speech Recognition, Identity Verification, Autonomous Driving, Natural Language Processing, Supervised Learning, Machine Translation, OCR Technology, Computer Vision, Machine Vision, Robot, Intelligent Question Answering, Expert System, Neural Network, Learning Algorithm, Automatic Reasoning, Driverless Technology, Mobile Internet, Industrial Internet, Digital Technology, Nano Computing, Intelligent Planning, Intelligent Optimization, Smart Wearables, Intelligent Manufacturing, Intelligent Customer Service, Intelligent Marketing, Digital Marketing, Unmanned Retail, Unmanned Factory, Mobile Payment, Third-Party Payment, NFC Payment, Human-Computer Interaction, Smart Agriculture, Intelligent Transportation, Intelligent Healthcare, Smart Home, Intelligent Investment Advisor, Intelligent Culture and Tourism, Intelligent Environmental Protection, Smart Grid, Intelligent Energy, Internet Healthcare, Internet Finance, Digital Finance, Fintech, Financial Technology, Quantitative Finance |
| AI asset | **Fixed assets:** Electronic equipment, data equipment, automation equipment, information equipment, server, intelligent terminal, computer room, communication equipment, integrated equipment, storage equipment, computing power, computer, chip, CPU, network |
|  | **Intangible assets:** Software, intelligence, information platform, system, data, digital, client, service platform, Internet, cloud computing, information technology, 5G, AI, Internet of Things, blockchain, technology, APP, mini-program, webpage, website |
| AI talents | Robot, Clustering algorithm, WEKA, Image processing, Scikit-learn, Unsupervised learning, Deep learning, Object recognition, Knowledge fusion, TensorFlow, Image enhancement, Image generation, NLP (Natural Language Processing), Dimensionality reduction, Facial expression recognition, Caffe, Transfer learning, Deeplearning4j, Recommender system, Semantic segmentation, Action recognition, Natural Language Processing, DL (Deep Learning), Character recognition, OpenCV, LDA (Latent Dirichlet Allocation), Knowledge extraction, Machine vision, Image sensor, Iris recognition, Neural network, HMM (Hidden Markov Model), Voice control, PyTorch, Boosting, Ensemble learning, Speech recognition, Traditional machine learning, Behavior monitoring, Image recognition, Deep Learning, Semantic processing, CNN (Convolutional Neural Network), Text recognition, Biometric recognition, MXNet, Speech enhancement, Libsvm, RF (Random Forest), Image detection, Image synthesis, Face recognition, Information extraction, Computational linguistics, Decision tree, Naive |

Bayes, Object recognition, SVM (Support Vector Machine), Word2Vec, OpenNLP, AR (Augmented Reality), Reinforcement learning, Semi-supervised learning, RNN (Recurrent Neural Network), Boosting, Ant colony algorithm, Computer vision, K-Means, Semantic retrieval, Drone, Fingerprint recognition, Image filtering, Word segmentation, Machine learning, Object tracking, Text mining, Image retrieval, SGD (Stochastic Gradient Descent), LSTM (Long Short-Term Memory), Image understanding, Machine intelligence, OCR (Optical Character Recognition), Classification algorithm, Image reconstruction, Semantic understanding, Gesture recognition, Dlib, Keras, Computer vision, Sentiment classification, Torch, Syntax analysis, Gesture interaction, Semantic analysis, Chatbot, Humanoid robot, Reinforcement learning, NLTK (Natural Language Toolkit), Semantic classification, Theano, Welding robot, Neural Networks, Syntax analysis, Matrix decomposition, LSA (Latent Semantic Analysis), Sentiment analysis, Motion capture, Image recognition, Speech synthesis, Visual intelligence, Markov model, Intelligent speech, Image matching, Bayesian network, Random Forest, Lexical analysis, Natural language processing, Xgboost, Pattern classification, Sound recognition, Voiceprint recognition, Inspection robot, Image calibration, Image classification, Liveness detection, Pattern recognition, Mahout, Virtual reality, Naive Bayes, Semantic recognition, Supervised learning,  Smart city, Smart card, Smart tablet, Smart manufacturing, Smart water management, Smart security, Autonomous driving, Smart lighting, Smart classroom, Smart transportation, Smart parking, AI healthcare, Smart terminals, Smart hospital, Smart scenic spots, Intelligent customer service, Smart doors, Smart meetings, Smart healthcare, Smart logistics, Smart assistance, Smart parks, Smart video, Smart elderly care, Intelligent machines, Smart agriculture, Smart teaching, Smart transportation, Smart energy, Smart auditing, Intelligent analysis, Smart government, Smart health, Smart tourism, Intelligent security, Smart party building, Smart education, Smart classroom, Smart teaching, Intelligent systems, Smart parking, Smart court, Smart community, Smart construction site, Smart judicial work, Smart voice, Smart environmental protection, Smart political and legal system, Smart operations, Smart industry, Intelligent control, Intelligent electronics, Smart urban management, Smart campus, Smart fire protection

# Online Appendix B. Additional test for domain AI readiness and AI capabilities

We further explored the relationship between a firm's domain AI readiness and its AI capabilities by adding firm fixed effects to the basis of Table 2. We found that the results in Columns 1 and 3 became insignificant, while the regression coefficients in Columns 2 and 4 remained significant at least at the 5% level. The results indicate that, in general, a firm's AI capabilities increase with the growth of AI readiness, and this increase is mainly reflected in the growth of AI talent recruitment.

Table A2 AI readiness and AI capability within firms

|  | (1) AI strategy | (2) AI talents | (3) AI asset | (4) AI capability |
|---|---|---|---|---|
| AI readiness (external technological evolvement) | -0.0055 (0.0077) | 0.0455** (0.0212) | 0.0654 (0.0760) | 0.9775*** (0.0172) |
| Log (Total assets) | 0.0271*** (0.0055) | 0.0507*** (0.0169) | 0.6712*** (0.0835) | 0.1185*** (0.0128) |
| Leverage | -0.0180 (0.0176) | 0.0349 (0.0498) | 0.5001** (0.2088) | -0.0011 (0.0383) |
| Tobin'q | 0.0012 (0.0017) | 0.0296*** (0.0068) | -0.0265 (0.0207) | -0.0007 (0.0039) |
| Firm FE | Yes | Yes | Yes | Yes |
| Year FE | Yes | Yes | Yes | Yes |
| # of Obs | 20,202 | 20,202 | 20,199 | 20,199 |
| $R^2$ | 0.5306 | 0.5779 | 0.8206 | 0.7562 |

# Online Appendix C. Robustness test for complementarity test

We sorted AI patents based on the absolute number of occurrences of IPC4 with the aim of intuitively identifying the domains where AI innovation is active. We recognize that such a measurement may lead to mismeasurement in fields where patents are not frequently filed. Therefore, we used the ratio of the number of occurrences of IPC4 in AI patents to the number of occurrences of IPC4 in all patents as the basis for sorting IPC4, and re-sorted IPC4 into deciles on an annual basis. Moreover, we calculated AI readiness based on this sorting method and conducted regression analysis. The regression results in Table A3 show that the results of AI readiness calculated using this alternative method are robust.

Table A3 An alternative measure of AI readiness

|  | (1) AI capability | (2) Lead. Revenue per employee | (3) Lead. TFP |
|---|---|---|---|
| AI readiness | 1.2756*** | | |
|  | (0.0290) | | |
| AI readiness* AI capability | | 0.0842*** | 0.0513*** |
|  | | (0.0284) | (0.0263) |
| Controls | Yes | Yes | Yes |
| Frim FE | No | Yes | Yes |
| Year FE | Yes | Yes | Yes |
| # of Obs | 20,412 | 16,417 | 15,850 |
| $R^2$ | 0.399 | 0.892 | 0.882 |

*Note*: The calculation of AI readiness in this table is based on the decile ranking of the ratio of the number of occurrences of IPC4 in AI patents to that in all patents, and the calculation process is similar to that in Table 4.

A firm's AI readiness may be endogenously related to its AI capabilities. Firms in domains with a high degree of AI readiness are more likely to develop AI capabilities, but firms with strong AI capabilities may also file more AI-related patents, thereby increasing their measured readiness. Therefore, the positive effects observed in Table 3 may be driven by firms in AI-intensive industries. As a robustness check, we excluded listed companies that have applied for AI patents and re-estimated the interaction effect between AI readiness and AI capabilities on the remaining sample. The estimated results are shown in Table A4, which indicate that the interaction effect between AI readiness and AI capabilities does not stem from those firms whose core business is directly related to AI.

Table A4 Excluding enterprises that have applied for AI patents

|  | (1) | (2) | (3) | (4) | (5) | (6) |
|---|---|---|---|---|---|---|
|  | Lead. Revenue per employee | | | Lead. TFP | | |
| AI readiness | 0.0023 | | | 0.0380 | | |
|  | (0.0026) | | | (0.0243) | | |
| AI readiness (external technological evolution) | | 0.0291 | | | 0.0399 | |
|  | | (0.0261) | | | (0.0243) | |
| AI readiness (internal strategic pivot) | | | -0.0000 | | | -0.0000 |
|  | | | (0.0001) | | | (0.0001) |
| AI capability | -0.0867*** | -0.0908*** | -0.0450** | -0.0727** | -0.0747*** | -0.0279 |
|  | (0.0291) | (0.0291) | (0.0229) | (0.0245) | (0.0245) | (0.0199) |
| AI readiness×AI capability | **0.0914***** | | | **0.0605**** | | |
|  | **(0.0352)** | | | **(0.0294)** | | |
| AI readiness (external technological evolution) ×AI capability | | **0.0954***** | | | **0.0641**** | |
|  | | **(0.0355)** | | | **(0.0294)** | |
| AI readiness (internal strategic pivot) ×AI capability | | | **0.0218** | | | **-0.0006** |
|  | | | **(0.0353)** | | | **(0.0297)** |
| Controls | Yes | Yes | Yes | Yes | Yes | Yes |
| Firm FE | Yes | Yes | Yes | Yes | Yes | Yes |
| Year FE | Yes | Yes | Yes | Yes | Yes | Yes |
| # of Obs | 10,777 | 10,777 | 10,777 | 10,343 | 10,343 | 10,343 |
| $R^2$ | 0.8936 | 0.8936 | 0.8935 | 0.9018 | 0.9018 | 0.9017 |

*Note*: The sample used in this table does not include companies that filed AI patents during the sample period.

# Online Appendix D. Exclusivity test of Instrumental Variable

To provide support for the validity of our instrument, we check whether the instrumental variables predict the future levels of productivity in companies that have not invested in AI following Martin & Yurukoglu (2017). The logic is as follows: if the instrument is valid, then it should be correlated only to the focal company's productivity through its effect on the focal company's AI capability. For companies without AI investments, we should not observe a significant correlation between the instrument and the outcome variable. To test this, we regress the dependent variable on the instrumental variables directly, using only data from observations that did not have AI talent hiring and AI asset.

The results in Table A5 show that, among the samples where both AI talent recruitment and AI assets are zero, the coefficient of the interaction term between AI policy and AI readiness is much smaller than that in the samples where at least one of the two is greater than zero, and all are insignificant. These results provide support for the validity of our instrumental variables.

Table A5 Exclusivity test of instrumental variables

| | (1) | (2) | (3) | (4) |
|---|---|---|---|---|
| Sample: | AI talent or AI asset >0 | AI talent, AI asset =0 | AI talent or AI asset >0 | AI talent, AI asset =0 |
| Dependent variable: | Lead. Revenue per employee | | Lead. TFP | |
| AI policy× AI readiness | 0.0376*** | -0.0040 | 0.0381*** | 0.0092 |
| | (0.0119) | (0.0171) | (0.0113) | (0.0170) |
| Controls | Yes | Yes | Yes | Yes |
| Firm FE | Yes | Yes | Yes | Yes |
| Year FE | Yes | Yes | Yes | Yes |
| # of Obs | 12,525 | 4,029 | 12,229 | 3,935 |
| $R^2$ | 0.8945 | 0.9165 | 0.9032 | 0.9227 |

*Note*:

# Online Appendix E. The consequences of adopting AI prematurely

Given that AlphaGo's victory over Lee Sedol in the game of Go in 2016 led to that year being regarded as the "first year of AI", we consider companies that recruited AI talent in 2016 and the subsequent year of 2017 as early adopters of AI after the first year of AI. In contrast, those that did not recruit AI talent during these two years are classified as late adopters. We conducted regression analyses to test the interaction effects separately within these two subsamples.

Table A5 presents the results of the subsample tests, and we find that the regression coefficients of the interaction terms for companies that adopted AI late are several times those of companies that adopted AI early. Meanwhile, the coefficient of the interaction term is significant only in the subsample of companies that adopted AI late. This indicates that if enterprises adopt AI before the arrival of the peak of AI readiness, they will not be able to obtain the benefits that AI readiness brings to AI adoption.

Table A6 A comparison between companies that adopted AI early and those late

|  | (1) | (2) | (3) | (4) |
|---|---|---|---|---|
| Sample: | Early adopter | Late adopter | Early adopter | Late adopter |
| Dependent variable: | Lead. Revenue per employee | | Lead. TFP | |
| AI readiness | -0.0830** | 0.0191 | -0.0489 | 0.0179 |
|  | (0.0390) | (0.0252) | (0.0435) | (0.0234) |
| AI capability | -0.0298 | -0.0790*** | -0.0255 | -0.0639*** |
|  | (0.0436) | (0.0281) | (0.0435) | (0.0234) |
| AI capability× AI readiness | 0.0125 | 0.1059*** | 0.0028 | 0.0741** |
|  | (0.0494) | (0.0339) | (0.0488) | (0.0301) |
| Controls | Yes | Yes | Yes | Yes |
| Firm FE | Yes | Yes | Yes | Yes |
| Year FE | Yes | Yes | Yes | Yes |
| # of Obs | 3,750 | 12,667 | 3,675 | 12,175 |
| $R^2$ | 0.8979 | 0.8899 | 0.9197 | 0.9018 |

# Online Appendix F. Calculation of Interdisciplinary AI Paper Growth

In this subsection, we outline how we computed the number of interdisciplinary AI papers. First, we extracted over 20 million academic papers published between 2016 and 2022 from the OpenAlex database, which aggregates data from Web of Science, Scopus, and other sources. Using the same AI lexicon as in the main text (Babina et al., 2024; Alekseeva et al., 2021), we matched papers' keywords and abstracts against this lexicon. Papers containing any AI-related terms were labeled as AI-related.

OpenAlex assigns hierarchical Concepts to papers, organized in a tree structure with 19 root-level categories and three descendant layers.[12] Approximately 85% of papers are tagged with at least one Concept. We counted Concept occurrences regardless of hierarchy, calculating growth as the difference between 2022 and 2016 counts in AI-related papers. After excluding AI-specific Concepts, Table A7 presents the top ten non-AI Concepts by growth.

Notably, Medicine ranked first, followed by Physics, Mathematics, Biology, and Psychology—all root-level Concepts. Only Medicine and Biology had corresponding IPC4 codes, while Physics, Mathematics, and Psychology did not. Among the 6-10th ranked Concepts, four were related to healthcare or biology. This confirms that biological and medical fields experienced the most significant growth in AI integration from 2016-2022.

Finally, we selected IPC4 codes G16B and G16H, which capture the intersection of healthcare/biology with algorithms, for the analysis in Section 5.2.

Table A7 The top 10 non-AI concepts in AI-related papers

| Rank | Concept | The hierarchy of concepts | The growth in the number of papers |
|---|---|---|---|
| 1 | **Medicine** | 1 | 72329 |
| 2 | Physics | 1 | 61993 |
| 3 | Mathematics | 1 | 60727 |
| 4 | **Biology** | 1 | 57202 |
| 5 | Psychology | 1 | 53853 |
| 6 | **Gene** | 2 | 52469 |
| 7 | **Neuro science** | 2 | 42422 |
| 8 | Quantum mechanics | 2 | 40740 |

---

[12] For specific information on the hierarchy of concepts in OpenAlex, please refer to: https://docs.openalex.org/api-entities/concepts.

| 9 | **Disease** | 2 | 38528 |
| 10 | **Cancer** | 2 | 35700 |